\documentclass[]{spie}
\usepackage[]{graphicx, xspace}
\usepackage[]{epstopdf}

\title{South Pole Telescope Software Systems: Control, Monitoring, and Data Acquisition}

\newcommand{\dfmux}{DfMux\xspace}
\newcommand{\kicp}{a}
\newcommand{\ucphys}{b}
\newcommand{\cardiff}{c}
\newcommand{\uc}{d}
\newcommand{\ucboulder}{e}
\newcommand{\nist}{f}
\newcommand{\mcgill}{g}
\newcommand{\efi}{h}
\newcommand{\argonneHEP}{i}
\newcommand{\ucAstro}{j}
\newcommand{\argonneMatSci}{k}
\newcommand{\berkeley}{l}
\newcommand{\ucboulderPhys}{m}
\newcommand{\caltech}{n}
\newcommand{\arizona}{o}
\newcommand{\michigan}{p}
\newcommand{\case}{q}
\newcommand{\minnesota}{r}
\newcommand{\saic}{s}
\newcommand{\harvard}{t}

\author{
K.~Story\supit{\kicp,\ucphys},
E.~Leitch\supit{\kicp},
P.~Ade\supit{\cardiff},
K.A.~Aird\supit{\uc},
J.E.~Austermann\supit{\ucboulder},
J.A.~Beall\supit{\nist},
D.~Becker\supit{\nist},
A.N.~Bender\supit{\mcgill},
B.A.~Benson\supit{\kicp,\efi},
L.E.~Bleem\supit{\kicp,\ucphys},
J.~Britton\supit{\nist},
J.E.~Carlstrom\supit{\kicp,\ucphys,\efi,\argonneHEP,\ucAstro},
C.L.~Chang\supit{\kicp,\efi,\argonneHEP},
H.C.~Chiang\supit{\kicp,\efi},
H-M.~Cho\supit{\nist},
T.M.~Crawford\supit{\kicp,\ucAstro},
A.T.~Crites\supit{\kicp,\ucAstro},
A.~Datesman\supit{\argonneMatSci},
T.~de Haan\supit{\mcgill},
M.A.~Dobbs\supit{\mcgill},
W.~Everett\supit{\kicp},
A.~Ewall-Wice\supit{\kicp,\ucphys},
E.M.~George\supit{\berkeley},
N.W.~Halverson\supit{\ucboulder,\ucboulderPhys},
N.~Harrington\supit{\berkeley},
J.W.~Henning\supit{\ucboulder},
G.C.~Hilton\supit{\nist},
W.L.~Holzapfel\supit{\berkeley},
S.~Hoover\supit{\kicp,\ucphys},
N.~Huang\supit{\kicp,\efi},
J.~Hubmayr\supit{\nist},
K.D.~Irwin\supit{\nist},
M.~Karfunkle\supit{\kicp,\ucphys},
R.~Keisler\supit{\kicp,\ucphys,\efi},
J.~Kennedy\supit{\mcgill},
A.T.~Lee\supit{\berkeley},
D.~Li\supit{\nist},
M.~Lueker\supit{\caltech},
D.P.~Marrone\supit{\arizona},
J.J.~McMahon\supit{\michigan},
J.~Mehl\supit{\kicp,\efi},
S.S.~Meyer\supit{\kicp,\ucphys,\efi,\ucAstro},
J.~Montgomery\supit{\kicp,\ucphys},
T.E.~Montroy\supit{\case},
J.~Nagy\supit{\case},
T.~Natoli\supit{\kicp,\ucphys},
J.P.~Nibarger\supit{\nist},
M.D.~Niemack\supit{\nist},
V.~Novosad\supit{\argonneMatSci},
S.~Padin\supit{\kicp},
C.~Pryke\supit{\minnesota},
C.L.~Reichardt\supit{\berkeley},
J.E.~Ruhl\supit{\case},
B.R.~Saliwanchik\supit{\case},
J.T.~Sayre\supit{\case},
K.K.~Schaffer\supit{\saic},
E.~Shirokoff\supit{\caltech},
G.~Smecher\supit{\mcgill},
B.~Stalder\supit{\harvard},
C.~Tucker\supit{\cardiff},
K.~Vanderlinde\supit{\mcgill},
J.D.~Vieira\supit{\caltech},
G.~Wang\supit{\argonneHEP},
R.~Williamson\supit{\kicp,\efi},
V.~Yefremenko\supit{\argonneHEP,\argonneMatSci},
K.W.~Yoon\supit{\nist},
E.~Young\supit{\berkeley}
\skiplinehalf
\supit{\kicp} Kavli Institute for Cosmological Physics, The University of Chicago, 5640 South Ellis Avenue, Chicago, IL 60637, USA\\
\supit{\ucphys} Department of Physics, University of Chicago, 5640 South Ellis Avenue, Chicago, IL, USA 60637 \\
\supit{\cardiff} Cardiff School of Physics and Astronomy, Cardiff University, Cardiff, United Kingdom\\
\supit{\uc} University of Chicago, 5640 South Ellis Avenue, Chicago, IL, USA 60637\\
\supit{\ucboulder} Department of Astrophysical and Planetary Sciences, University of Colorado, Boulder, Colorado,80309, USA\\
\supit{\nist} NIST, Boulder, CO 80305, USA\\
\supit{\mcgill} McGill University, Montreal, Quebec, Canada\\
\supit{\efi} Enrico Fermi Institute, University of Chicago, 5640 South Ellis Avenue, Chicago, IL, USA 60637\\
\supit{\argonneHEP} High Energy Physics Division, Argonne National Laboratory, Argonne, IL 60439, USA\\
\supit{\ucAstro} Department of Astronomy and Astrophysics, University of Chicago, 5640 South Ellis Avenue, Chicago, IL, USA 60637\\
\supit{\argonneMatSci} Materials Science Division, Argonne National Laboratory, Argonne, IL 60439, USA\\
\supit{\berkeley} University of California, Berkeley, 151 LeConte Hall Berkeley, CA 94720, USA\\
\supit{\ucboulderPhys} Department of Physics, University of Colorado, Boulder, CO 80309\\
\supit{\caltech} California Institute of Technology, Pasadena, CA 91125, USA\\
\supit{\arizona} Steward Observatory, University of Arizona, 933 North Cherry Avenue, Tucson, AZ 85721, USA\\
\supit{\michigan} University of Michigan, Ann Arbor, Michigan, USA\\
\supit{\case} Case Western Reserve University, Cleveland, Ohio 44106, USA\\
\supit{\minnesota} University of Minnesota, Minneapolis, MN 55455, USA\\
\supit{\saic} School of the Art Institute of Chicago, Chicago, Illinois, 60603, USA\\
\supit{\harvard} Harvard-Smithsonian Center for Astrophysics, 60 Garden Street, Cambridge, MA, USA 02138\\
}

\authorinfo{Further author information: (Send correspondence to K.Story)\\K.Story: E-mail: kstory@uchicago.edu}

\begin{document}
\maketitle

\begin{abstract}
We present the software system used to control and operate the South Pole Telescope. The South Pole Telescope is a 10-meter millimeter-wavelength telescope designed to measure anisotropies in the cosmic microwave background (CMB) at arcminute angular resolution.  In the austral summer of 2011/12, the SPT was equipped with a new polarization-sensitive camera, which consists of 1536 transition-edge sensor bolometers.  The bolometers are read out using 36 independent digital frequency multiplexing (\dfmux) readout boards, each with its own embedded processors. These autonomous boards control and read out data from the focal plane with on-board software and firmware.  An overall control software system running on a separate control computer controls the \dfmux boards, the cryostat and all other aspects of telescope operation.  This control software collects and monitors data in real-time, and stores the data to disk for transfer to the United States for analysis.
\end{abstract}


\section{Introduction}
The South Pole Telescope (SPT) is a 10-meter off-axis Gregorian telescope located at the NSF Amundsen-Scott South Pole Station in Antarctica \cite{carlstrom11}.  
Construction on the SPT was finished in 2007, and from 2008-2011 the SPT completed a survey of $\sim$2500 deg$^2$ of sky at observing frequencies of 95, 150, and 220 GHz.
In this original survey, the SPT was used to measure the millimeter-wavelength sky in temperature at frequencies: 95, 150 and 220 GHz.  

In the austral summer of 2011/12, the SPT was equipped with a new polarization-sensitive camera, a combination that we will hereafter refer to as SPTpol.  The primary science goal of SPTpol is to measure the polarization of the cosmic microwave background (CMB).  
The SPTpol camera consists of two types of pixels sensitive to either 90 or 150 GHz light\cite{chang12, hubmayr11}.  In total there are 768 pixels, 180 at 90 GHz and 588 at 150 GHz, with each pixel consisting of two bolometers that are sensitive to orthogonal polarization modes, for a total of 1536 detectors. Each pixel has a pair of detectors which are sensitive to orthogonal polarization modes.  We use transition-edge sensor bolometers that operate at a temperature of $\sim 500$ mK\cite{irwin98a, lee98}.

In this paper we describe the software system that controls, monitors and acquires data from SPTpol.

\subsection{Science Goals}
The primary goal of SPTpol is to measure anisotropies in the polarization of the CMB on angular scales of a few degrees to a few arcmintues\cite{austermann12}.  
These measurements can be used to constrain several cosmological parameters, including the sum of the neutrino masses and the energy scale of Inflation, the theorized exponential expansion of the universe that occured shortly after the Big Bang.
The SPTpol measurement of CMB polarization anisotropy can be decomposed into even-parity (E-mode) and odd-parity (B-mode) modes.  We expect that SPTpol will make the best measurement to date of the E-modes, and could make the first-ever detection of B-modes.
On scales of several arcminutes, B-modes are generated by the gravitational lensing of the CMB by large-scale structure, and are sensitive to the sum of the neutrino masses through the effect of massive neutrinos on the growth of structure\cite{zaldarriaga98}.  On scales of a few degrees, B-modes are predicted to be generated by primordial gravity waves from Inflation, and their detection are generally considering both evidence for the theory of Inflation and the energy scale of the physics associated with it\cite{seljak97}.

\subsection{Detector Control and Readout Hardware}
\label{subsection:dataReadoutHardware}

The details of the detector readout hardware are important for understanding the software system of SPTpol, and we give an overview of the readout hardware here.

The SPTpol camera consists of 1536 transition-edge sensor bolometers that operate at $\sim 500$ mK.  
To read out this many detectors while maintaining the low operating temperature is thermally challenging due to the wiring heat-load on the refrigerator. 
In SPTpol, this problem is alleviated using a digital frequency multiplexing (\dfmux) scheme, where a set of 12 detectors are read out on a common pair of wires\cite{dobbs11, smecher12a}.
The modulated signal from each set of 12 detectors is amplified by a Superconducting Quantum Interference Device\cite{dobbs11} (SQUID) operating at a temperature of $\sim 4$ K, followed by a low-noise amplifier.  The signal is then digitized and demodulated using FPGA-based warm readout electronics, which we refer to as \dfmux boards \cite{smecher12a}.

\begin{figure}
   \centering
   \resizebox{0.6\linewidth}{!}{\includegraphics{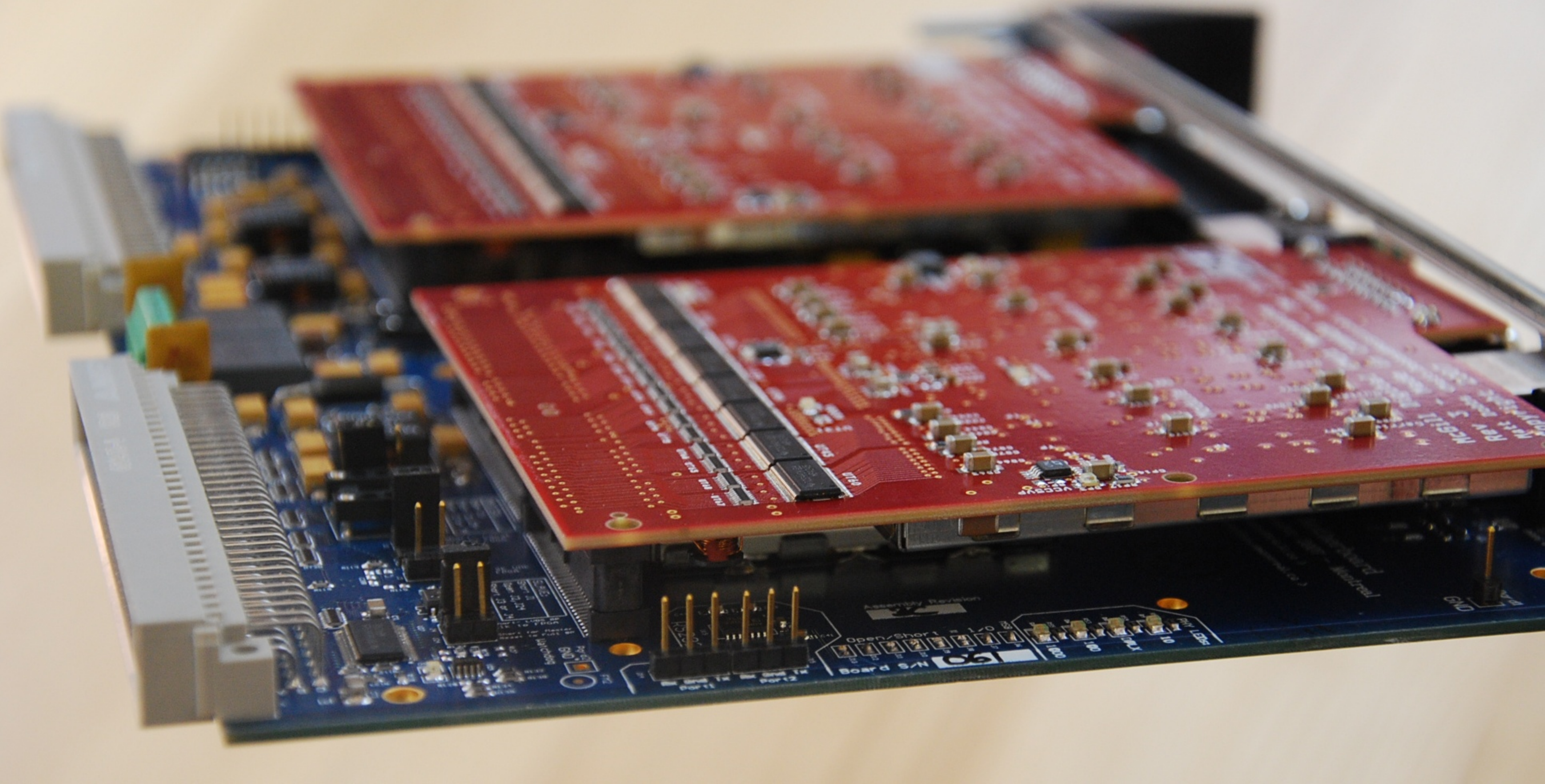}}
      \caption{A \dfmux board, which provides the hardware and software interface between the detectors and the rest of the telescope system.  These boards are $\sim 6 \times 10$ inches, and run autonomous processors that control and read data from the detectors. [Photo Credit: Adam Gilbert]}
   \label{fig:dfmuxboard}
\end{figure}

\dfmux boards (see Figure \ref{fig:dfmuxboard}) provide the hardware and software interface between the detectors and the rest of the telescope system.  SPTpol uses 36 \dfmux boards to read out and control its detectors, and each \dfmux board is responsible for controlling and operating four SQUIDs and 48 detectors.  The \dfmux boards have independent Xilinx MicroBlaze$^{\hbox{\tiny{TM}}}$ processors running Linux, and can run autonomously, receiving commands from the telescope control system and returning data  via an Ethernet connection.

\subsection{Refrigerator Control and Monitoring Hardware}

The refrigerator which cools the detectors and cold readout electronics is controlled and monitored by three independent cryogenic refrigerator control boards (hereafter referred to as cryoelectronics boards).  The cryoelectronics boards are designed in a similar way to the \dfmux boards and share much of the on-board software stack\cite{smecher12b}.  Each cryoelectronics board controls up to eight resistive heaters in the refrigerator, and reads out up to sixteen thermometers.


\section{The Software Control System: the GCP}

\begin{figure}
   \centering
   \resizebox{0.9\linewidth}{!}{\includegraphics{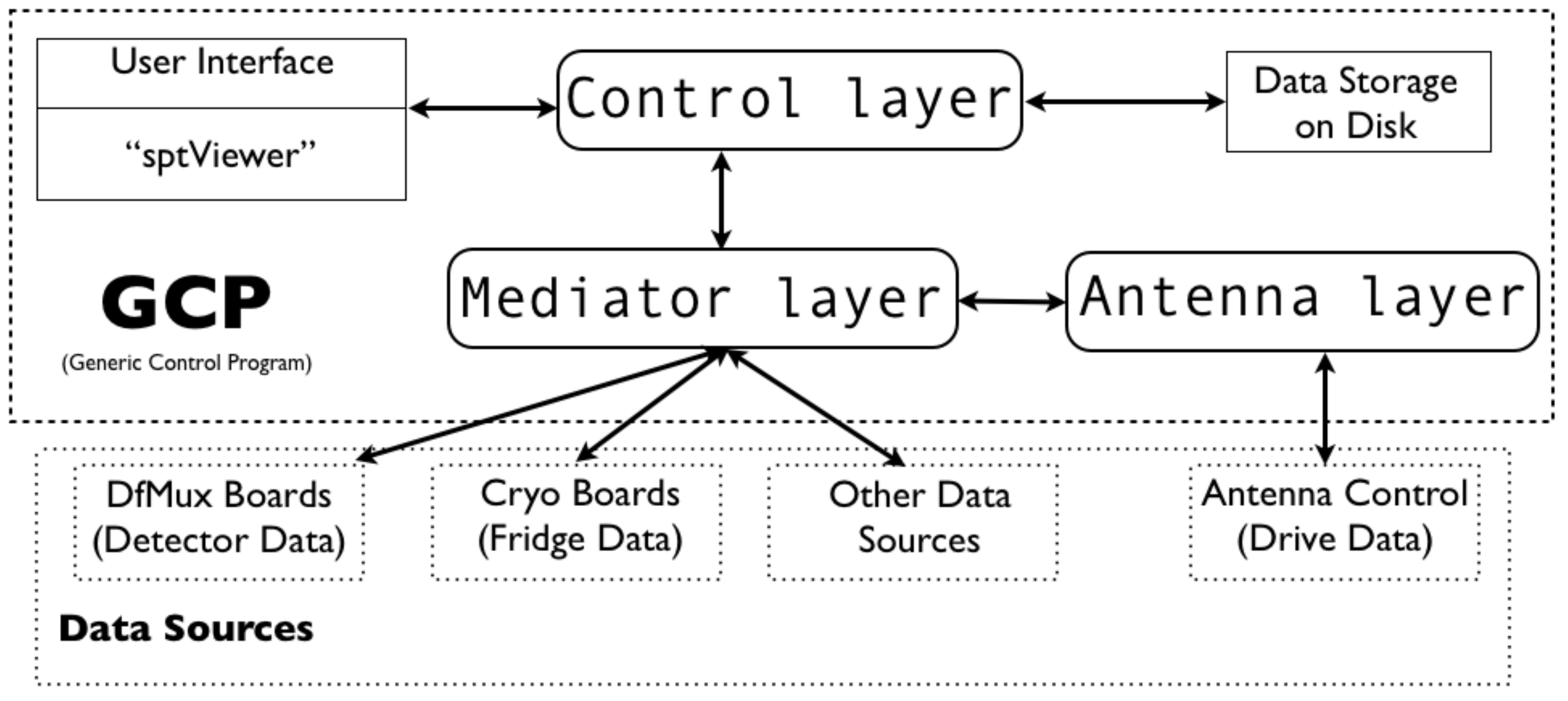}}
   \caption{The layout of the GCP software system.  The main components of the GCP software are labeled in the upper box, and examples of data sources are labeled in the lower box.}
   \label{fig:gcplayout}
\end{figure}

SPTpol is run by a self-contained software system called the Generic Control Program (GCP).  The GCP runs at all times, controlling and operating all aspects of the experiment.  The GCP is primarily written in the C++ and C programming languages, with a few processes in the Python language and a Tcl/Tk-based graphical user interface.  A schematic layout of the GCP software is shown in Figure \ref{fig:gcplayout}.

The original version of the GCP was developed by Erik Leitch for the Sunyaev-Zel'dovich Array experiment in 2003-04 \cite{muchovej07}, modeled after the control system developed by Martin Shepherd for the Cosmic Background Imager\cite{padin01}.  Core components of the graphical interface (sptViewer, discussed below) and scheduling language were adopted from the CBI control system.  Since that time, the GCP has been modified to run nine additional experiments, most recently including SPTpol, the Keck-array \cite{sheehy10}, BICEP2 \cite{nguyen08}, and PolarBear \cite{arnold09}.

The GCP accomplishes several tasks.  First, it provides a robust way to control all aspects of the experiment.  This includes commanding the proprietary software and hardware that controls the movements of the telescope itself, managing the refrigerator, and operating the detectors.  Second, the GCP collects, collates and stores data from many different sources.  These sources include -- but are not limited to -- data from the detectors, auxiliary ``house-keeping'' data that record the condition of the detectors, data that record the bearing positions and the state of the drive system, temperatures from the refrigerator, and weather conditions.  Finally, the GCP provides real-time monitoring of all aspects of the telescope.  This is critical for smooth and efficient operation of the experiment.

The GCP must be stable and use computer resources efficiently, since it runs the entire experiment and operates at all times.   For the SPTpol system, the GCP components typically use less than $45\%$ of a single 2.13~GHz core and well below the available 16 GB of RAM in the SPTpol system.  For SPTpol, the GCP has run for timescales of a month between restarts, while the original version of the GCP for the SZA ran for up to eight months between restarts.

The GCP is designed to be modular and light-weight.  All GCP processes communicate through simple TCP socket connections, which makes it easy to adapt the system architecture to project-specific requirements; processes can be distributed among multiple machines, as needed to meet the CPU-loading or hardware requirements.  Socket communications are implemented using low-level Unix libraries, making data transfer efficient and allowing for greater portability between platforms.  All components of the GCP control system support asynchronous reconnection, which allows components to be re-started individually without affecting the entire system.  The GCP code uses only a small number of open-source third-party libraries, avoiding licensing restrictions and minimizing platform-dependence. The GCP has been compiled and run on a wide variety of 2.4 and 2.6 Linux OS kernels as well as Mac OS 10.4---10.7, with typical $\sim2$~GHz single-core compile times of $\sim5$~minutes.

The GCP control software has three primary components: the Control layer, the Mediator layer, and the Antenna layer.  The top-level Control layer provides overall control of the entire experiment, as well as monitoring and archiving of all data. The graphical user interface communicates with the Control layer for control and monitoring. The Mediator layer mediates between the top-level Control layer and all low-level hardware-control processes.  The purpose of the Mediator layer is two-fold: to forward commands to all underlying processes, and to collate data from different sources (potentially sampled at different rates) throughout the system.  These data  are repackaged into monolithic time-stamped ``data frames," which are archived by the Control layer.
The final software layer is the Antenna layer, which controls the telescope mount.  For SPTpol, the Antenna layer provides an interface between the GCP and two proprietary Vertex computers that drive the telescope: the Antenna Control Unit (ACU) and the Secondary Control Unit (SCU).  The ACU provides low-level control of the drive motors and readout of encoders and power supplies, while the SCU handles motion control of the cryostat that holds the cooled secondary mirror and focal plane.  Commands from the Control layer are passed to the ACU and SCU via the Antenna layer.

\section{User Interface}

Users interact with the SPTpol control software primarily through a graphical user interface (GUI), called the ``sptViewer,'' which communicates with the Control layer via a bi-directional socket connection.  The queue of programs for controlling the telescope (hereafter referred to as ``schedules'') is managed from the sptViewer interface, which also displays messages logged from all GCP control processes.  Finally, the sptViewer provides the primary interface for graphical monitoring of data from the experiment.  All data sources are available for real-time monitoring, with built-in support for user-configurable plotting and text displays, real-time power spectrum analysis, and built-in widgets for common tasks, like control and monitoring of frame grabbers for optical pointing.

\begin{figure}
   \centering
   \resizebox{0.80\linewidth}{!}{\includegraphics{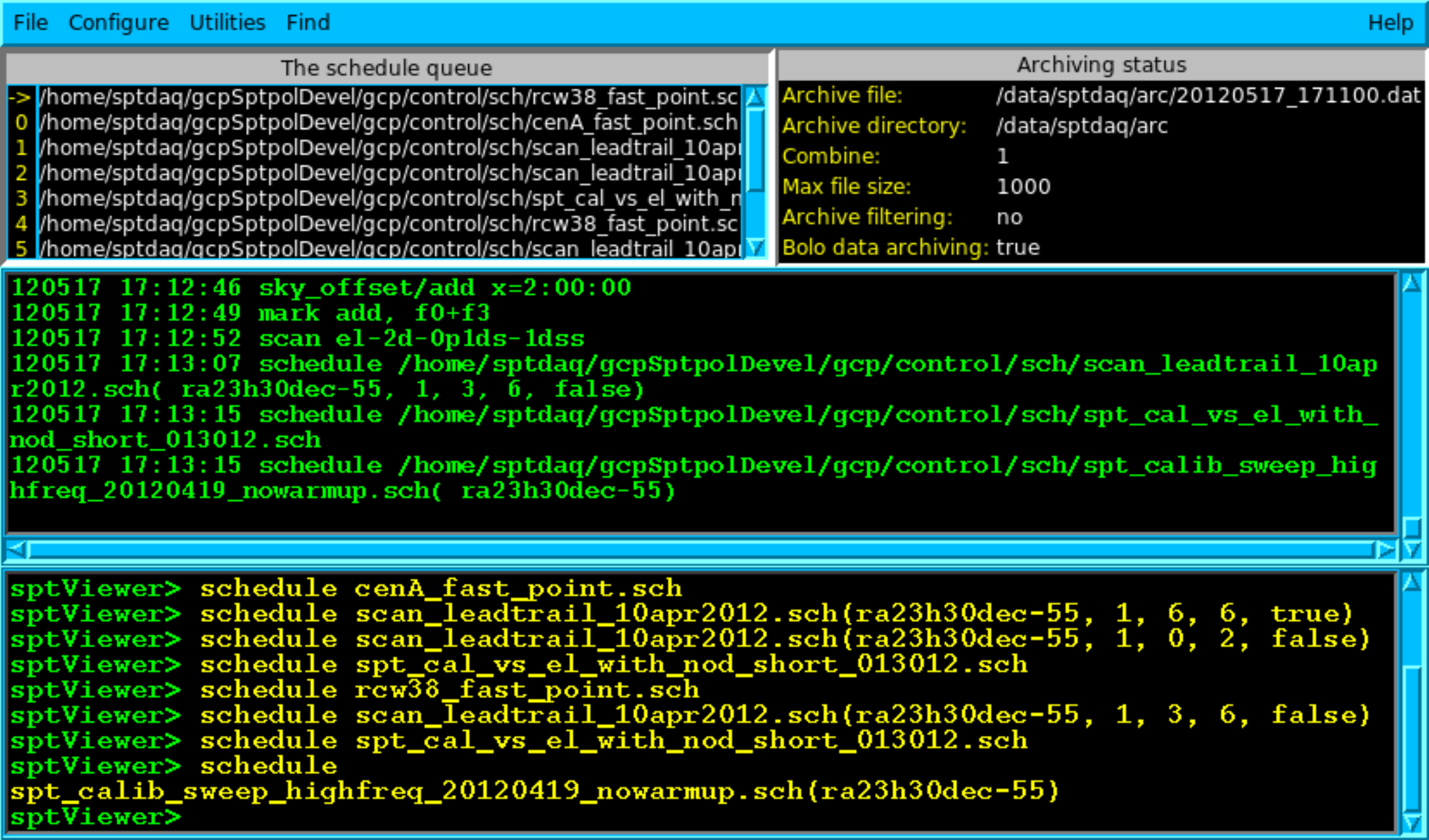}}
   \caption{The sptViewer, the main user interface to the GCP.  Schedules and commands are entered in the bottom panel, the middle panel displays the log messages being generated by the control layer, and the top panel lists the current queue of schedules on the left, and the status of data archiving on the right.}
   \label{fig:sptViewer}
\end{figure}

\subsection{Drive Controls and GCP Schedules}
\label{sec:DCandGS}
The ACU has exclusive control of the telescope drive system, and can be commanded in three ways: manually from the ACU interface, manually from the command line in the sptViewer interface to the GCP, and automatically from GCP schedules.  The manual ACU interface is only used for maintenance, debugging and one-time-only tasks.  All normal operations are mediated  by the GCP.

Schedules are written in a GCP-specific schedule language, parsed by an internal run-time interpreter, and perform all high-level tasks including scanning the telescope for science observations, cycling the refrigerator, observing astronomical calibration sources, and operating the internal calibration source.  All commands are defined in a single master schedule library, and individual schedules draw commands from this library.  Schedules are called from the command line in the sptViewer.  An example is shown in Figure \ref{fig:sptViewer}, where the most recently queued schedule is ``$spt\_calib\_sweep\_highfreq\_20120419\_nowarmup.sch$," which observes the internal calibration source.  Schedules are called in sequence, and the GCP puts sequential schedules into a queue, executing one after another.  In Figure \ref{fig:sptViewer}, the queue of schedules can be seen in the upper-left.  Manual commands issued from the sptViewer interface are simply interleaved with the currently executing schedule, allowing simultaneous reconfiguration or control of hardware systems that do not interfere with the running schedule.

The GCP controls the ACU via a serial connection from the the Antenna layer.  A PCI timing card generates a 100~Hz pulse synchronized to GPS time, which is delivered to the ACU and the Antenna layer via software interrupts.  This pulse serves as the fundamental clock for the tracking communication loop between the Antenna layer and the ACU (see \S\ref{sec:data_acquisition} for details).  The Antenna layer is isolated on a separate computer to ensure that other GCP or user-space processes cannot introduce latencies in the 100~Hz interrupt delivery.  In this way the entire GCP system can be deployed on off-the-shelf Linux boxes running a non-realtime Linux 2.6 kernel.

\subsection{Detector and Refrigerator Control Software}
\label{sec:DandFContlol}

The detector control software is comprised of a well documented body of code\cite{macdermid09, smecher12a, smecher12b} written in Python. This software is organized as a set of ``algorithms" that perform specific tasks related to the detectors and readout system, and control scripts which call the underlying algorithms.  Algorithms are run on the control computer and pass low level commands to the \dfmux boards (see \S\ref{subsection:dataReadoutHardware}).  Examples of algorithms are routines that control various heaters, tune SQUIDs, turn on detectors, and align timestamps.  Control scripts in Python encapsulate these algorithms into high-level function calls.  These high-level control scripts are then called by the GCP, through the GCP schedule language.

The efficiency of this software is greatly enhanced by using the Python multi-processing library.  Many processes can be made much more efficient by multi-threading.  For example, when tuning SQUIDs, it is much more efficient to run one algorithm per thread on each of the $36 \times 4=144$ SQUIDs, as opposed to calling the tuning algorithm on each SQUID sequentially.  An illustrative example of the benefits of this multi-threaded code is the time-stamp alignment of detector data.  The \dfmux boards must be told to align to specific ticks in a 25 MHz stream of timestamps.  When the \dfmux boards were commanded to ``align timestamps" simultaneously using multi-threading rather than sequentially, timestamp alignment improved from $\sim$1 millisecond to a few tens of nanoseconds, an improvement of more than four orders of magnitude.

The Mediator layer in the GCP calls the high-level control scripts, which send commands to the \dfmux boards over an Ethernet connection.  \dfmux boards then execute the commanded algorithms independently, and report back to the Mediator layer.  \dfmux boards are autonomous and integrated into the system in a modular way, so in the rare case when a \dfmux board crashes only the data from its detectors are lost.  All aspects of direct detector control are managed by the \dfmux boards \cite{smecher12a}.

The refrigerator control software is very similar to the detector control software.  It follows the same model as the detector readout, with algorithms that are called by high-level control scripts, which can in turn be called from the GCP.  Algorithms are then passed to the cryoelectronics boards, as described in \S\ref{subsection:dataReadoutHardware}.



\begin{figure}
   \centering
   \resizebox{0.59\linewidth}{!}{\includegraphics{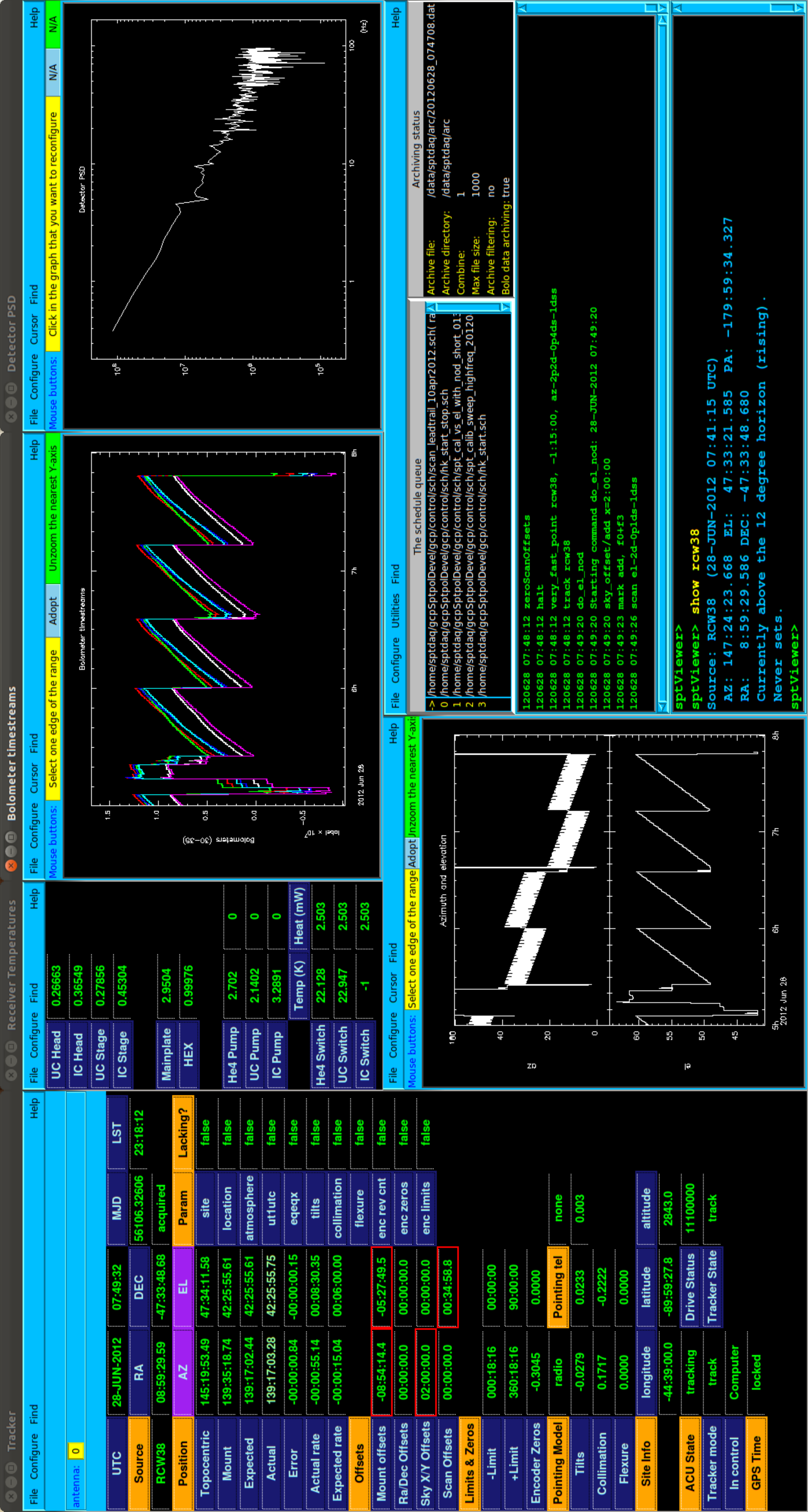}}
   \caption{In SPTpol, data are monitored primarily with the sptViewer.  This figure displays the real-time monitoring capabilities provided by the sptViewer software.  
     We describe each window starting from the left and moving clockwise.  The first window is the ``Tracker'' GUI page, which displays the encoder positions and other data related to the telescope mount.  Boxes in red show data whose valitidy conditions have been violated.  The ``Receiver Temperature'' GUI page displays the thermometer temperatures and heater power from the refrigerator.  The ``Bolometer timestreams'' window plots detector timestreams in real-time.  The ``Detector PSD'' window plots the raw power spectrum of a single detector.  The sptViewer window is described in Figure \ref{fig:sptViewer}.  The last plot shows multi-pannel plotting, in this case plotting the real-time pointing of the telescope during a science observation.}
   \label{fig:TrackerPage}
\end{figure}

\section{Data Monitoring}
It is essential to monitor all aspects of the telescope in real-time.  This allows the user to monitor the health of the experiment and discover problems shortly after they arise.

In SPTpol, data are monitored in several ways, the first of which is real-time monitoring with the sptViewer.  To accomplish this, the GCP Control layer contains a server thread that listens for connection requests from sptViewer clients, and serves the requested portion of the current data frame over a socket connection.  SptViewer clients can request portions of data as large as the entire data frame, or as small as a single data register, allowing users to monitor live data even over a low-bandwidth connection, such as the satellite link to the South Pole.  Users can create, save and load GUI pages and plots from the sptViewer interface for data monitoring.  Examples of the monitoring capabilities provided by the sptViewer are shown in Figure \ref{fig:TrackerPage}.

All data registers can be assigned validity conditions in the Control layer; when these conditions are violated, the control system can generate a variety of alerts, including emails, pager activation,  and calls to LMT radios, which the telescope operators at the South Pole carry at all times.

The GCP processes include built-in support for logging messages to timestamped files on disk.  These logs are important for monitoring the ongoing progress of the experiment and back-tracing any problems that occur.  They are also used to determine ancillary information for data analysis, such as the start and stop times of particular schedules.

\begin{figure}
   \centering
   \resizebox{0.99\linewidth}{!}{\includegraphics{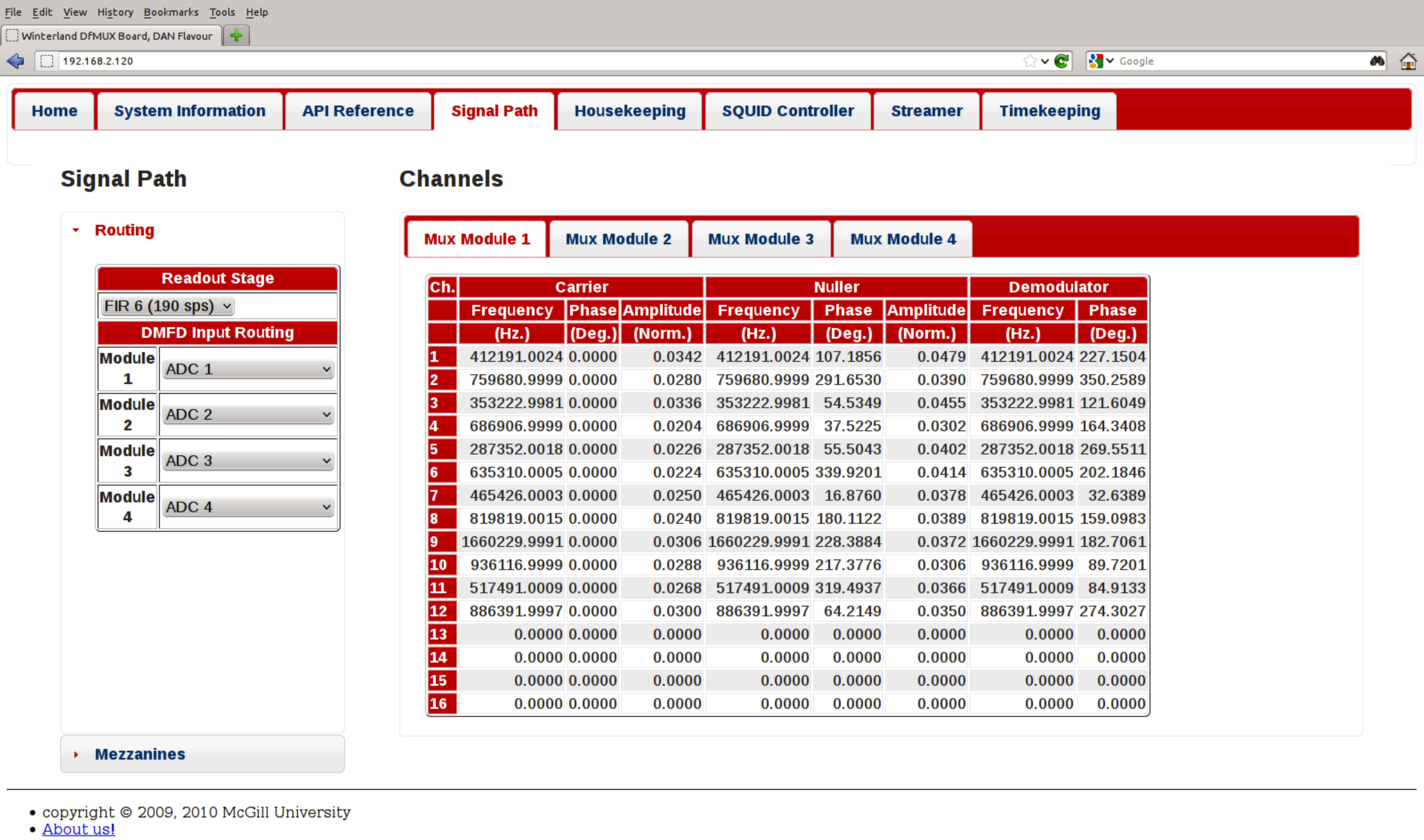}}
   \caption{The web interface for a \dfmux readout board.  This web interface provides both an interface to control the \dfmux board, as well as easy monitoring of \dfmux board processes.}
   \label{fig:dfmuxWebpage}
\end{figure}

The \dfmux and cryoelectronics boards have their own web interfaces which can be used to monitor all auxiliary information about the detectors and the refrigerator (see Figure \ref{fig:dfmuxWebpage}).  The power supplies which run the crates of \dfmux and cryoelectronics boards also have their own web interfaces, allowing for remote power cycling of the boards.

\section{Data Acquisition}
\label{sec:data_acquisition}

The SPTpol software system collects data from a wide variety of sources and processes, with data collated by the Mediator layer in the GCP.  The Mediator layer does this in software with C++ classes called ``Consumers."  Examples of Consumers include a BolometerConsumer for collecting data from the bolometer detectors, a HousekeepingConsumer for collecting auxiliary readout data, a CrythonConsumer for recording temperatures from the cryogenics, and an AntennaConsumer for gathering information about the telescope positions and drive currents.

The fundamental unit of data in the GCP is a 1-second ``data frame."  Each data frame represents the data from every monitored system in the experiment for that one second.  Individual datum entries are called ``data registers.'' The data from most sources do not change quickly (for example, the temperatures in the refrigerator), and are recorded at the 1-Hz frame rate.  For all of the 1-Hz data, the Mediator layer asks its Consumers to poll their various data sources for new values, and the Consumers asynchronously returns values to the Mediator for packing into the data frame.

The two notable exceptions to this model are telescope pointing data and detector data.  Telescope pointing data are collected by the Antenna layer at 100~Hz rate from the serial interface to the ACU, and are then passed to the Mediator layer through the Mediator's AntennaConsumer class.
Bolometer detector timestreams are sampled at a rate of 25 MHz.  These data are down-sampled by a factor of $2^{17}$ in the \dfmux boards, which returns the data at a rate of 25 MHz / $2^{17} \approx 190.7$ Hz.  The telescope pointing data are interpolated to match the detector data rate in post-processing.

Data from each detector are read out by a DfMux board and parsed into UDP packets. The data packets are pushed through an ethernet connection to a separate software layer on the control computer called the ``DfMuxMonitor.''  The detector data from all detectors must be grouped with micro-second accuracy.  To achieve this, each detector data sample is assigned a GPS-derived timestamp at the time when when the detector signal is demodulated in the \dfmux board.  Data packets from all \dfmux boards are unpacked and matched based on timestamps by the DfMuxMonitor software.  When data arrives from all \dfmux boards for a given timestamp (corresponding to a single 190.7 Hz interval), the DfMuxMonitor passes the complete set of detector data to the BolometerConsumer in the Mediator layer, where the detector data are packed into the data frame.  In the SPTpol system, timestamps between different \dfmux boards are aligned to better than a microsecond, and most are aligned to better than the timestamp resolution, which is ten nano-seconds.  In a normal observation, the GCP does not drop any bolometer data.

Auxiliary data from the \dfmux boards, and all data from the cryoelectronics boards are collected through Python interface layers.  Two independent Python servers run in the background, transferring data from the boards to the GCP.  These servers poll the boards for data, however the boards are not always able to return data at a rate of 1 Hz.  The servers keep a buffer of the most recently acquired data, and when the GCP polls the Python server for new data, the server returns the values in the buffer.

As discussed in \S\ref{sec:DCandGS}, a timing card in the antenna computer generates the 100~Hz pulse which drives the tracking loop.  The same card outputs a 1 pulse-per-second signal and IRIG-B timecode which is delivered to dedicated timing boards in the three crates of \dfmux and cryoelectronics boards.
A master timing board receives the timing signals and passes them to two timing boards that are daisy-chained to the master.
Each timing board sends out a 25~MHz signal across the backplane of the crate to all \dfmux and cryoelectronics boards in its crate.  
These timestamps are then synchronized as described above, and are used to match data in the GCP.

\subsection{Archiving Data}

Data frames are created and filled in the Mediator layer, then passed to the Control layer for archiving.  Data are archived to disk in the form of binary archive files.  In the GCP code, a user-defined structure call an ``arraymap'' defines a three-tiered hierarchy of data items that constitute the data frame.  The archive file contains a header, which is a binary representation of the arraymap, followed by a configurable number of data frames, each written as a byte array containing all the data for a single frame.  The GCP codebase provides C++ libraries for reading archived data, and built-in support for compiling archive readers for IDL, Matlab, and Python.  Data are stored at the South Pole on RAID arrays, and are sent back to the United States via satellite each day.  The total compressed data takes up $\sim 60$ Gigabytes per day.

\section{Summary / Conclusion}
The SPTpol telescope software system is a multi-faceted body of software which manages all aspects of the SPTpol telescope control, data monitoring and acquisition.  
The core of this software is formed by a modular, light-weight software system called the GCP.
The GCP controls all aspects of the telescope and provides the primary user interface to the system.  
The detector readout system in SPTpol relies on autonomous \dfmux boards, which control, operate and monitor the detectors and interface with the GCP.
The GCP also provides real-time monitoring of all aspects of operation and data acquisition on the telescope.  

\acknowledgements{
The South Pole Telescope is supported by the National Science Foundation through grants ANT-0638937 and ANT-0130612. Partial support is also provided by the NSF Physics Frontier Center grant PHY-0114422 to the Kavli Institute of Cosmological Physics at the University of Chicago, the Kavli Foundation and the Gordon and Betty Moore Foundation. 
The McGill authors acknowledge funding from the Natural Sciences and Engineering Research Council, Canadian Institute for Advanced Research, and Canada Research Chairs program. M. Dobbs acknowledges support from an Alfred P. Sloan Research Fellowship.''
Work at NIST is supported by the NIST Innovations in Measurement Science program. 
The work at Argonne National Laboratory, including the use of facility at the Center for Nanoscale Materials (CNM), was supported by Office of Science and Office of Basic Energy Sciences of the U.S. Department of Energy, under Contract No. DEAC02-06CH11357. 
Technical support from Nanofabrication Group at the CNM, Argonne National Laboratory, under User Proposal $\#$164 and $\#$467, is gratefully acknowledged.
}

\bibliography{sptpol_software.bib}
\bibliographystyle{spiebib}
\end{document}